\documentclass[12pt]{article}
\usepackage{latexsym}
\usepackage{amssymb, amsmath, xspace, lscape,  latexsym}

\renewcommand{\dim}{\mathrm{dim \,}}
\renewcommand{\mod}{\mbox{ mod }}

\renewcommand{\le}{\leqslant}
\renewcommand{\ge}{\geqslant}

\newcommand{\vf}{\varphi}

\def\beq#1#2\eeq{%
        \begin{equation}%
        \label{#1}%
            #2%
        \end{equation}%
    }

\newcommand{\lb}[1]{\label{#1}}

\newcommand{\mref}[1]{(\ref{#1})}

\newcommand{\p}{\partial}

\renewcommand{\a}{\alpha}

\newcommand{\R}{\mathbb R}

\newcommand{\C}{\mathbb C}

\newcommand{\Z}{\mathbb Z}

\newcommand{\anm}{{{\cal A}_n(m)}}

\newcommand{\cnm}{{{\cal C}_{n+1}(m,l)}}

\renewcommand{\hat}{\widehat}
\renewcommand{\tilde}{\widetilde}

\def\btheor#1\etheor{%
        \begin{theor}%
            #1%
        \end{theor}%
    }

    \def\bsled#1\esled{%
        \begin{sled}%
            #1%
        \end{sled}%
    }

\newcommand{\ol}[1]{\overline{#1}}
\newcommand{\cl}{{\cal L}}
\newcommand{\cR}{{\cal R}}
\newcommand{\cA}{{\cal A}}

\newtheorem{theorem}{Theorem}

\newtheorem{prop}{Proposition}
\newtheorem{cor}{Corollary}

\newcommand{\nad}[2]{\genfrac{}{}{0pt}{}{#1}{#2}}

\def\hm#1{#1\nobreak\discretionary{}{\hbox{\m@th$#1$}}{}}
\def\mi#1{\discretionary{\hbox{\m@th$#1$}}{\hbox{\m@th$#1$}}{}}

\vspace{4ex}
\begin{document}

\vspace*{1cm}

\begin{center}
{\Large \bf Quasi-invariants of dihedral systems}

\vspace{1cm}

{\bf  M. Feigin }

\end{center}

\vspace{5mm}

\noindent {A basis of quasi-invariant module over invariants is explicitly
constructed for the two-dimensional Coxeter systems with arbitrary multiplicities.
It is proved that this basis consists of $m$-harmonic polynomials, thus the
earlier results of Veselov and the author for the case of constant
multiplicity are generalized.}

\vspace{5mm}

%{\bf Keywords:} Quasi-invariants of
%Coxeter systems, $m$-harmonic polynomials, Calogero--Moser
%systems.

%{\bf Subjects:} 81R12, 20F55.

\section{Introduction}

We start with the introducing the main object of our
considerations, the algebra of quasi-invariants. Consider a Coxeter
system ${\cal R}$ in $\R^n$ which consists of the non-collinear pairs of
 vectors
$\{\pm\a\}$ with prescribed multiplicities $m_\a\in\Z_+$. The corresponding
 Coxeter group
$G$ is generated by the reflections
$s_\a$, $\a\in \cR$:
$$
s_\a u = u - \frac{2(\a,u)}{(\a,\a)}\a, \quad u \in \R^n.
$$
The group $G$ should be finite and the set of reflections $\{s_\a\}$
must be the set of all the reflections in $G$.
The multiplicity function $m_\a= m(\a)$ is supposed to be
invariant:
$$
m_\a=m_{g(\a)}
$$
for any $g\in G$, $\a\in\cR$.

A polynomial $p(x)$ is called  {\it quasi-invariant} related to the system
$\cR$ if  it satisfies the condition
\beq{quazi}
s_\a p(x) - p(x) = O( (\a,x)^{2m_\a+1} )
\eeq
near the hyperplanes $(\a,x)=0$ for any $\a\in\cR$.
Equivalently,
$$
\p_\a^{2s-1} p =0, \quad \mbox{if} \, \, (\a,x)=0, \quad  s=1,\ldots, m_\a.
$$
For instance, if a polynomial $p(x)$ belongs to the ring $S^G$
of  polynomials invariant under the geometric action of the Coxeter group,
then  $s_\a p(x) = p(x)$ for any $\a\in\cR$, and therefore property \mref{quazi}
is satisfied.
Thus the quasi-invariants $Q^\cR$
form a ring containing the ring of invariants $S^G$.

The rings of quasi-invariants were introduced by Chalykh and
Veselov in 1990 in the context of quantum integrable systems
\cite{CV}. The authors showed that to each quasi-invariant $q(x)$
it corresponds a differential operator $\cl_q=\chi(q)$ of the form
$$
\cl_q(x,\p_x)=q(\p_x)+ \,\mbox{lower order terms},
$$
and all such operators  $\cl_q$ commute.
Under the homomorphism $\chi$ the quasi-invariant $q(x)=x_1^2+\ldots+x_n^2$
corresponds to the generalized Calogero--Moser operator \cite{C3,M,OP2}
\beq{cmoperator}
\cl_{x^2}=\Delta-\sum_{\a\in\cR_+}\frac{2m_\a}{(\a,x)}\p_\a.
\eeq
Here $\cR_+$ is a subset of vectors from $\cR$ belonging to a
halfspace, and $\Delta$ is the Laplace operator in $\R^n$.
Therefore each quasi-invariant $q(x)$ leads to the quantum
integral $\cl_q$ of the Calogero--Moser problem \mref{cmoperator}.
As it is explained in
\cite{FV2} the commutative ring of all quantum integrals for
\mref{cmoperator} is isomorphic to the ring of quasi-invariants
$Q^\cR$.

We note that the homomorphism $\chi$ on the invariant
polynomials was considered by Heckman and Opdam for a general
multiplicity function \cite{HO}. It is crucial that for the integer
multiplicity function the ring $Q^\cR$ of quantum integrals becomes larger
than the ring of invariants $S^G$, and the ring depends on these integer
parameters, as it all was discovered by Chalykh and Veselov \cite{CV}.
The necessity to consider integer multiplicities to get larger commutative
rings have been justified recently by Taniguchi \cite{T}.

The first investigation of the ring $Q^\cR$ was done by Volchenko, Kozachko,
 and Mishachev
\cite{VKM} in the case of dihedral system with multiplicity equal~1.
The authors found the multiplicative generators of the ring.
The more complete description of the ring $Q^\cR$ for
the dihedral systems with constant multiplicity function $m_\a=m\in\Z_+$
was carried by Veselov and the author \cite{FV2}.
It turned out that for the dihedral system $I_2(N)$
consisting of $N$ lines on the plane
$\R^2\cong \C$ the ring $Q^{I_2(N)}$
is a free module over the subring of invariants $S^{I_2(N)}=\C[z \bar z,
z^N+\bar
z^N]$. These module has $2N$ generators $h_1, \ldots,h_{2N}$ such that the
linear space  $H_m=\langle h_1 \ldots,h_{2N}\rangle$ spanned by
these generators is the solution space of the following system of two
equations
$$
\cl_{z\bar z} h =0,
$$
$$
\cl_{z^N+\bar z^N} h =0,
$$
and the determinant formulas for $h_i$ were found.
It was conjectured in  \cite{FV1}
that the ring of quasi-invariants $Q^\cR$ is free as a module over
invariants $S^G$ for any Coxeter system $\cR$. It was also
conjectured that one can take $m$-harmonic polynomials as a basis
in this free module. The space $H_m$ of $m$-harmonic polynomials consists of
the solutions to the following system of equations
$$
\chi(\sigma) h=0,
$$
where $\sigma$ is an arbitrary element from the ring $S^G_+$ of
invariant polynomials with free term equal zero (see \cite{FV1}).

Etingof and Ginzburg remarkably proved in \cite{EG} that the ring $Q^\cR$  is
free over invariants $S^G$ for any Coxeter system $\cR$.
But the choice of a basis in  $H_m$ as a free basis for the quasi-invariant module is
impossible in general (see counterexample in \cite{EG}).
Nevertheless the degrees of homogeneous generators coincide with
the degrees of homogeneous basis in $H_m$, as it is proved in
\cite{EG}. The free generators should form a basis in the
complement to ideal generated by homogeneous invariants of
positive degree in $Q^\cR$. The problem to produce some particular choice
of such a basis remains open.

Besides the Cohen--Macaulay property for the
quasi-invariant algebra, the Gorenstein property was also
established by Etingof and Ginzburg \cite{EG}.
One of the two proofs given in \cite{EG} is based on the results of
Felder--Veselov paper \cite{FeV}. In that paper the authors in
particular compute the Poincare polynomial for the space $H_m$ of $m$-harmonic
polynomials. The Gorenstein property corresponds to the
palindromicity of the Poincare polynomial for $H_m$ which was
conjectured in \cite{FV1}.

The differential operators on the singular algebraic variety corresponding
to $Q^\cR$ were studied by Berest, Etingof, and Ginzburg in
\cite{BEG}. The authors proved that the algebra of differential
operators is simple and Morita equivalent to the Weyl algebra of
polynomial differential operators in $n$-dimensional space.

In the present paper following \cite{F} we describe the quasi-invariants of dihedral
systems
$I_2({2N})$
with arbitrary (invariant) multiplicities
$(m,n)$\footnote{In this case there exists two orbits
under the action of the group $G$ on the set of roots $\cR$.
The multiplicity of one of them equals $m\in\Z_+$, and it is equal
to $n\in\Z_+$ on another orbit.}.
We give explicit formulas for the generators of the ring $Q^{I_2(2N)}$
as a module over invariants $S^{I_2(2N)}$ generalizing the formulas from
\cite{FV2}. It occurs that in this
case as in the case of dihedral systems with constant multiplicity \cite{FV2},
one can take as the generators a homogeneous basis in the space
 $H_m$\footnote{We use notation
$H_m$ meaning by $m$ {\it the function} of multiplicity $m=m(\a)$.
For the system
$I_2(2N)$ the function of multiplicity takes two values $m$ and $n$.}
of $m$-harmonic
polynomials.
Therefore we obtain explicit determinant formulas for the basic
$m$-harmonic polynomials for arbitrary dihedral systems.

The technique used in this paper is different from the approach in
\cite{FV2} where dihedral systems with constant multiplicity were
considered. Now we heavily use the results of Etingof--Ginzburg
\cite{EG} and Felder--Veselov \cite{FeV}. This approach allows to
reduce the explicit calculations essentially. The scheme of our
considerations is the following.

In Section 2 we review the result of Felder and Veselov on the
Poincare polynomial for $m$-harmonics, and we write down explicitly the
Poincare polynomial $P_{H_m}$ for dihedral $m$-harmonics.
By Etingof--Ginzburg theorem the degrees of homogeneous generators
for quasi-invariants have degrees given by $P_{H_m}$. In Section 3
we prove that these generators can be chosen to have a particular
form. To do this we argue as follows. We take a general polynomial
of degree under consideration and impose quasi-invariant
conditions. From Etingof--Ginzburg theorem it follows  that
this system of equations must have a solution.
From the form of equations it follows that it must also exist
a solution of the desired form.
Then we check that the linear span of the quasi-invariants of the
form constructed does not belong to the ideal generated by
homogeneous invariants of positive degree. This can be easily
done using again the  Etingof--Ginzburg theorem. Thus
we conclude (by Etingof--Ginzburg theorem) that we have constructed a
free basis with elements of particular form.

In Section 4 we check firstly that the constructed basis is
$m$-harmonic. This is due to the special form of the chosen basis,
here we again use the Etingof--Ginzburg theorem.
Then we come back to the system of quasi-invariant conditions which defined
our basis. We note that for a polynomial of our form the set of
quasi-invariant conditions leads to $m$-harmonicity of this
polynomial. In particular, the polynomial of our form is uniquely
determined by the set of quasi-invariance equations. This fact
allows us to get explicit determinant formulas for $m$-harmonic
polynomials and therefore for the basis of quasi-invariant module.

%%%%%%%%%%%%%%%%%%%%%%%%

\section{Poincar\'e polynomial for dihedral groups}

Let $P_{H_m}(t)$ be Poincar\'e polynomial for the space $H_m$ of
$m$-harmonic polynomials related to a Coxeter system $\cR$,
$$
P_{H_m}(t) = \sum_{i\ge 0} \dim (H_m^{(i)}) t^i,
$$
where $H_m^{(i)}$ is the space of homogeneous $m$-harmonic
polynomials of degree $i$. The following formula for $P_{H_m}(t)$
was found by Felder and Veselov in \cite{FeV},
\beq{poincare}
P_{H_m}(t)=\sum_{V_j} t^{\sum_a m_a d_a^-(V_j)} P_j(t).
\eeq
In this formula the external sum is carried over all non-isomorphic irreducible
representations of a Coxeter group $G$. The internal sum is carried over
the
classes $C_a$ of conjugate reflections in $G$. Polynomial $P_j$ is
the Poincar\'e polynomial for representation $V_j$ in the space
$H_0 \cong \C[x_1, \ldots, x_n]/I_0$, where $I_0$ is the ideal in the ring of polynomials
generated by homogeneous invariants of positive degree. That is, the coefficient at $t^k$ equals
the dimension of isotypic component of the representation $V_j$
in the $k$th grade of
$\C[x_1, \ldots, x_n]/I_0$.
Then
$$
d_a^-(V_j)=\frac{2N_a \dim V^-_{j,a}}{\dim V_j},
$$
where $N_a$ is a number of elements in the class $C_a$, and
$$
V_{j,a}^- = \{v\in V_j | s_\a v = -v \mbox{ for any } s_\a\in C_a
\},
$$
$m_a$ is a multiplicity for reflections $s_\a\in C_a$.

For dihedral group $I_2(N)$ with odd $N$ there are two
one-dimensional non-isomorphic irreducible representations and
$\frac{N-1}2$ two-dimensional ones. All the reflections are conjugate
and $N_a=N$.
Since in the complex coordinates the ring of invariants is isomorphic to $\C[z\bar
z, z^N+\bar z^N]$, one can choose generators in factor-space as
follows
$$
\C[z, \bar z]/I_0 \simeq \langle 1, z, \bar z, \ldots,
z^{N-1}, \bar z^{N-1}, z^N-\bar z^N\rangle.
$$
For one-dimensional representations the piece of sum
(\ref{poincare}) has the form
$$
1+t^{(2m+1)N}
$$
where the first term corresponds to the trivial representation
$V_{triv}$ which is realized on the constants. The second term
corresponds to the sign representation $V_{sign}$ which is realized
on the vector $z^N-\bar z^N$. We've also used that
$$
d_a^-(V_{triv})=\frac{2N \cdot 0}1=0,
$$
$$
d_a^-(V_{sign})=\frac{2N\cdot 1}1=2N.
$$
Consider now two-dimensional representations. Representation $V_i$
is realized as $\langle z^i,\bar z^i\rangle$ and as $\langle z^{N-i},\bar
z^{N-i}\rangle$. Then
$$
d_a^-(V_i) = \frac{2N\cdot 1}2=N.
$$
Totally we get
$$
P_{H_m}=1+t^{(2m+1)N}+\sum_{i=1}^{\frac{N-1}2} 2t^{mN}
(t^i+t^{N-i}) = 1+2\sum_{i=1}^{N-1} t^{mN+i} + t^{(2m+1)N}.
$$

Now let us consider even dihedral system $I_2(2N)$. We
have two classes $C_a$ and $C_b$ of conjugate reflections each of which consists
of $N$ elements. Denote the multiplicities of reflections as $m_a=m$ and $m_b=n$.
As in the odd case
$$
\C[z, \bar z]/I_0 \simeq \langle 1, z, \bar z, \ldots,
z^{2N-1}, \bar z^{2N-1}, z^{2N}-\bar z^{2N}\rangle.
$$
At first we analyze one-dimensional representations and calculate
the corresponding terms in the Poincar\'e polynomial. There are four
one-dimensional representations for $I_2(2N)$ depending on whether
each class of reflections $C_a, C_b$ acts as 1 or -1.
\begin{itemize}
\item
$V_{triv}=\langle 1\rangle$. The corresponding term in
(\ref{poincare}) is equal to 1.
\item
$V_{sign}=\langle z^{2N}-\bar z^{2N}\rangle$,
$d_a^-(V_{sign})=d_b^-(V_{sign})=\frac{2N\cdot 1}1=2N$.
The contribution to the Poincar\'e polynomial is $t^{(m+n+1)2N}$.
\item
$V_{sign^1}=\langle z^N+\bar z^N\rangle$. One gets
$$
m_a d_a^-(V_{sign^1})+ m_b d_b^-(V_{sign^1}) = m\cdot \frac{2N\cdot 0}1 +
n\cdot \frac{2N\cdot 1}1= 2nN
$$
which leads to $t^{(2n+1)N}$ term in (\ref{poincare}).
\item
$V_{sign^2}=\langle z^N-\bar z^N\rangle$. One gets
$$
m_a d_a^-(V_{sign^2})+ m_b d_b^-(V_{sign^2}) = m\cdot \frac{2N\cdot 1}1 +
n\cdot \frac{2N\cdot 0}1= 2mN
$$
which leads to $t^{(2m+1)N}$ term in (\ref{poincare}).
\end{itemize}

Indeed, representations $V_{sign^1}, V_{sign^2}$ are defined by
the property that for any $s\in C_a, \tau\in C_b$
$$
s|_{V_{sign^1}} = Id, \quad s|_{V_{sign^2}} = -Id,
$$
$$
\tau|_{V_{sign^1}} = -Id, \quad \tau|_{V_{sign^2}} = Id.
$$
The dihedral group is generated by two reflections $s\in
C_a, \tau\in C_b$ such that $(s\tau)^{2N}=1$. Also it is generated by
$s$ and $s\tau$. In our notations we can think of $s$ as a
reflection with respect to line $z=\bar z$, and $s\tau$ would be a
rotation by the angle $\phi=\frac{\pi}N$. Thus the following
formulas hold
\beq{sst}
\begin{gathered}
s: z\to\bar z,\,\bar z\to z,
\\
s\tau: z\to\epsilon z,\, \bar z\to \bar\epsilon\bar z,
\end{gathered}
\eeq
where $\epsilon=e^{i\frac{\pi}{N}}$.
Then representation $V_{sign^1}$ is characterized  by the property
$$
s|_{V_{sign^1}} = Id,
\tau|_{V_{sign^1}} = -Id,
$$
or equivalently
$$
s|_{V_{sign^1}} = Id,
(s\tau)|_{V_{sign^1}} = -Id.
$$
Obviously the following action of group elements holds
$$
s(z^N+\bar z^N) = \bar z^N + z^N,
$$
and
$$
(s\tau)(z^N+\bar z^N) = \epsilon^N z^N +\bar \epsilon^N\bar z^N =
-(z^N+\bar z^N).
$$
Thus $V_{sign^1}=\langle z^N+\bar z^N\rangle$ and if $m_s=m, m_\tau=n$ then the contribution of
this irreducible representation to the Poincar\'e polynomial would
be $t^{(2n+1)N}$. Analogously $V_{sign^2} = \langle z^N-\bar
z^N\rangle$ since
$$
s(z^N-\bar z^N) = \bar z^N - z^N = -(z^N-\bar z^N),
$$
and
$$
(s\tau)(z^N-\bar z^N) = \epsilon^N z^N -\bar \epsilon^N\bar z^N =
-(z^N-\bar z^N),
$$
the corresponding term in the Poincar\'e polynomial is $t^{(2m+1)N}$.

Now let's turn to two-dimensional irreducible representations.
Representation $V_i$ is realized twice in
$\C[z, \bar z]/I_0$ as the space
$\langle z^i,\bar z^i\rangle$ and as $\langle z^{2N-i},\bar
z^{2N-i}\rangle$, $i=1,\ldots,N-1$. Thus $P_i(t)=2t^i+2t^{2N-i}$ in
(\ref{poincare}). For all the reflections $s\in C_a, \tau\in C_b$
one has $\dim {V_i^-} = 1$, hence
$d_a^-(V_i)=d_b^-(V_i)=\frac{2N\cdot 1}2=N$
and we get the term in the Poincar\'e polynomial equal to
$2t^{(m+n)N}(t^i+t^{2N-i})$.
Summing up
\beq{poindi}
P_{H_m}= 1 + t^{(m+n+1)2N}+  t^{(2m+1)N} + t^{(2n+1)N} +
2\sum_{i=1}^{N-1} t^{(m+n)N}( t^i + t^{2N-i}).
\eeq

\section{Generators of quasi-invariants}

We recall the following key theorem proven by Etingof and
Ginzburg \cite{EG} (see also \cite{FV1}).
\begin{theorem}\cite{EG}\label{EGt}
For any Coxeter system ${\cR}$ the ring of quasi-invariants
$Q^\cR$ is free as a module over invariants $S^G$ of the
corresponding Coxeter group $G$.
As the generators one can take any homogeneous representatives of
the factor space $Q^\cR/I$, where
$I$ is the ideal in $Q^\cR$ generated by the
homogeneous invariant polynomials of positive degrees. The degrees
of the homogeneous generators coincide with the degrees of a homogeneous
basis in the space $H_m$ of $m$-harmonic polynomials.
\end{theorem}
 We are going to determine what are the
polynomials that can be taken as the generators for dihedral groups.
As the case of
constant multiplicity function is already considered in \cite{FV1}
 we shall suppose that
our group is even dihedral group $I_2(2N)$ with different multiplicities, for instance $m>n$.
By Theorem~\ref{EGt} the degrees of generators for $Q^{I_2(2N)}$
over invariants are same as the degrees of basic $m$-harmonic
polynomials. According to (\ref{poindi}) they satisfy
$$
(2n+1)N<(m+n)N+i<(2m+1)N<(m+n+1)2N,
$$
where $1\le i\le 2N-1, i\ne N$. And we get the following table for
number of generators of corresponding degree:
$$
\begin{array}{ccccccc}\label{table}
\deg : & & 0 &  (2n+1)N & (m+n)N+i & (2m+1)N & (m+n+1)2N \nonumber\\
&&&&&\\
\nad{\mbox{number of}}{\mbox{generators}}:& & 1 & 1 & 2 & 1 & 1\nonumber\\
\end{array}
$$
Let us introduce the normal vectors $\a_i, i=0,\ldots,2N-1$ to the lines of
reflections. Namely, $\a_i=(-\sin\frac{\pi i}{2N}, \cos\frac{\pi
i}{2N})$, the multiplicities $m_i=m$ for even $i$ and $m_i=n$ for
odd $i$.
Let's define the following four quasi-invariants
\begin{eqnarray}
q^0=1,\,\, q^1=(z^N+\bar z^N)^{2n+1} = \mu_1
\prod_{\nad{i=1;}{i=2 j + 1}}^{2N-1}(\a_i,x)^{2n+1},\nonumber\\
q^2=(z^N-\bar z^N)^{2m+1} = \mu_2
\prod_{\nad{i=0;}{i=2j}}^{2N-2}(\a_i,x)^{2m+1},\label{q}\\
q^3=(z^N+\bar z^N)^{2n+1}(z^N-\bar z^N)^{2m+1} = \mu_3
\prod_{i=0}^{2N-1}(\a_i,x)^{2m_i+1}\nonumber,
\end{eqnarray}
where $\mu_1, \mu_2, \mu_3$ are some constants.
%where summation in $q^1$ is done for odd indexes $i$, summation in
%$q^2$ is for even indexes $i$, and summation in $q^3$ is carried
%for all $i$.
It's obvious that $q^0$ and $q^3$ are quasi-invariants. Polynomial
$q^1$ is also quasi-invariant. Indeed, conditions on the lines
$(\a_i,x)=0$ with odd $i$ are obviously satisfied. If $i$ is even
then the conditions follow from the invariance
$s_{\a_i}q^1=q^1$. Analogously, $q^2\in Q^{I_2(2N)}$.

These quasi-invariants will be a part of the basis for $Q^{I_2(2N)}$ over
invariants that we are constructing.

\begin{prop}
Quasi-invariants $ q^0, q^1, q^2, q^3$ defined by (\ref{q}) do not
belong to the ideal $I$ generated by homogeneous invariants of positive
degree $S^{I_2(2N)}_+$ in the ring $Q^{I_2(2N)}$.
\end{prop}

\noindent
{\bf Proof }
According to Theorem \ref{EGt} the degrees of free generators
for $Q^{I_2(2N)}$ over invariants are given by the table on page
\pageref{table}. Therefore there are no quasi-invariants of degree
less than $\deg q^1$ which are not invariants.
Since $q^1$ is not invariant we conclude that $q^1\notin I$. Let us show that $q^2\notin I$.
Let $r^{1,2}_i$ be independent elements in the complement to $I$
of degrees $(m+n)N+i$, $1\le i \le 2N-1, i\ne N$.
Suppose that $q_2\in I$, that is
\beq{q2}
q^2=s_0 q^0+s_1 q^1+\sum_i s^1_i r^1_i +\sum_i s^2_i r^2_i,
\eeq
where $s_0, s_1, s^1_i, s^2_i$ are invariants, i.e. they belong to
$\C[z\bar z, z^{2N}+\bar z^{2N}]$. Since $\deg r^{1,2}_i \ne
0 \mod N$ polynomials $s_i^{1,2}$ are divisible by $z\bar z$. Let
us consider relation (\ref{q2}) modulo terms divisible by $z\bar
z$. We get
$$
z^{(2m+1)N}-\bar z^{(2m+1)N} = \lambda_1(z^{2N}+\bar z^{2N})^a
(z^{(2n+1)N}+\bar z^{(2n+1)N})+ \lambda_2(z^{2N}+\bar z^{2N})^b +
O(z\bar z).
$$
Now we get a contradiction as it follows both
$\lambda_1+\lambda_2=1$, and $\lambda_1+\lambda_2=-1$. Therefore
relation (\ref{q2}) is impossible and $q^2\notin I$.

Let us finally show that $q^3\notin I$. As above we obtain that
the following relation must hold
$$
z^{2(m+n+1)N} - \bar z^{2(m+n+1)N} = \lambda_1 (z^{2N}+\bar
z^{2N})^a \left(z^{(2m+1)N}-\bar z^{(2m+1)N}\right)+
$$
\beq{q3}
 \lambda_2 (z^{2N}+\bar
z^{2N})^b \left(z^{(2n+1)N}+\bar z^{(2n+1)N}\right) +\lambda_3(z^{2N}+\bar
z^{2N})^c +O(z\bar z).
\eeq
We obtain $\lambda_1=\lambda_2=0$ as for any $a,b$ the degrees of
monomials in $z$ and in $\bar z$ which are obtained from the first
two terms of the right hand-side in (\ref{q3}) have the form
$N \cdot(\mbox{odd number})$. And in the left-hand side the result of
dividing degree by $N$ is $2(m+n+1)$ which is even. Further we
conclude that the right-hand side of (\ref{q3}) takes the form
$\lambda_3(z^{2Nc}+\bar z^{2Nc}) +O(z\bar z)$, and the equality
(\ref{q3}) is impossible. Hence $q^3\notin I$ and the proposition is
proved.

The proved proposition shows that polynomials $q^0,q^1,q^2,q^3$ may be chosen
as a part of free basis for $Q^{I_2(2N)}$ over invariants. The rest $2N-4$
polynomials in a basis consist of pairs of polynomials of degrees
$(m+n)N+i$ with $1\le i \le 2N-1, i\ne N$. The next Proposition and Theorem show
that these quasi-invariants may be chosen to have more or less simple form.

\begin{prop}\label{freebasis}
%As a free basis for $Q_m$ over $S^{I_2(2N)}$ one can take
%quasi-invariants $q^0, \ldots, q^3$ defined by $(\ref{q})$ and
%quasi-invariants $q_i^{1,2}$ of the following form
There exist quasi-invariants $q_i^{1,2}$ of the form
\beq{bazis}
\begin{gathered}
q^1_i = \sum_{s=0}^{m+n} a_{s} z^{(m+n-s)N+i} \bar z^{Ns}, \quad
a_0=1,
\\
q^2_i =\sum_{s=0}^{m+n} b_{s}\bar z^{(m+n-s)N+i} z^{Ns}, \quad
b_{s} = \bar a_{s}
\end{gathered}
\eeq
for $1\le i \le 2N-1, \, i\ne N$.
\end{prop}

\noindent
{\bf Proof }
%According to the proposition above we are left to
%prove that there are quasi-invariants of the form $q^1_i, q^2_i$
%and that they do not belong to ideal $I_m$. The first part is more
%difficult and we start with it.
Let $r^{1,2}_i$ be some basis in the homogeneous complement to
$I$ of degree $(m+n)N+i$, $1\le i \le 2N-1, i\ne N$. Let us
consider two-dimensional space $\langle r^1_i, r^2_i\rangle$. As it
does not intersect $I$ we may think that
$$
r^1_i=z^{(m+n)N+i} +
O(z\bar z),
$$
$$
r^2_i=\bar z^{(m+n)N+i} + O(z\bar z).
$$
%More than this, if there exists a pair of quasi-invariants of the form
%as above then they do not belong to ideal $I_m$ and altogether
%with (\ref{q}) they define a free basis for $Q_m$ over
%$S^{I_2(2N)}$.
Let us consider the quasi-invariance conditions
for the first polynomial
$$
r^1_i=\sum_{s=0}^{(m+n)N+i-1} a_s
z^{(m+n)N+i-s}\bar z^s, \quad a_0=1.
$$
The lines $(\a_j,x)=0$,
$j=0,\ldots,2N-1$ are given by the equations $z=\epsilon^{j}\bar
z$ where $\epsilon^{2N}=1$, and $\epsilon$ is a primitive root.
The quasi-invariance condition $\p^{2t-1}_{\a_j} r^1_i =0$ at
$(\a_j,x)=0 \Leftrightarrow z=\epsilon^j \bar z$ takes the form $$
\sum_{s=0}^{(m+n)N+i-1} a_s
((m+n)N+i-2s)^{2t-1}\epsilon^{((m+n)N+i-s)j}= 0, $$ or, equivalently,
$$
\sum_{s=0}^{(m+n)N+i-1} a_s
\left((m+n)N+i-2s\right)^{2t-1}\left(\frac1{\epsilon^s}\right)^j= 0.
$$
Let $t$ satisfy $1\le t\le n< m$. Then quasi-invariance conditions are
nontrivial for all $2N$ lines $j=0,\ldots 2N-1$. The conditions
may be rewritten in the form
$$
\sum_{p=0}^{2N-1} A_p \left(\frac1{\epsilon^p}\right)^j= 0,
$$
where
$$
A_p= \sum_{s\equiv p (2N)} a_s
((m+n)N+i-2s)^{2t-1}
$$
This system of equations is of Vandermonde type.
As $\epsilon^{p_1}\ne\epsilon^{p_2}$ if $p_1\ne p_2$ we conclude
that
$$
A_p=0, \quad p=0,\ldots, 2N-1.
$$
Now let us consider the left quasi-invariance conditions
corresponding to $n<t\le m$. These conditions correspond to the
lines with even $j$. We have
$$
\sum_{s=0}^{(m+n)N+i-1} a_s
\left((m+n)N+i-2s\right)^{2t-1}\left(\frac1{\epsilon^s}\right)^j= 0,
$$
where $j=0,2,\ldots,2N-2$. Equivalently,
$$
\sum_{p=0}^{N-1} B_p \left(\frac1{\epsilon^{2p}}\right)^j= 0,
$$
where
$$
B_p= \sum_{s\equiv p (N)} a_s
((m+n)N+i-2s)^{2t-1},
$$
and $j=0,1,\ldots, N-1$.
This system of equations is again of Vandermonde type.
As $\epsilon^{2p_1}\ne\epsilon^{2p_2}$ if $p_1\ne p_2$ and $0\le p_i\le N-1$ we conclude
that
$$
B_p=0, \quad p=0,\ldots, N-1.
$$
Thus quasi-invariance of the polynomial
$$
r^1_i=\sum_{s=0}^{(m+n)N+i-1} a_s
z^{(m+n)N+i-s}\bar z^s
$$
is equivalent to the following conditions on its coefficients
\beq{qcond}
\begin{gathered}
 \sum_{s\equiv p (2N)} a_s
((m+n)N+i-2s)^{2t-1} =0, \quad p=0,1,\ldots, 2N-1, \,\, 1\le t \le n,
\\
 \sum_{s\equiv p (N)} a_s
((m+n)N+i-2s)^{2t-1} =0, \quad p=0,1,\ldots, N-1, \,\, n < t \le m.
\end{gathered}
\eeq
These equations are split to the systems of equations for the
sets of coefficients $\{a_s\}$ with indexes $s$ having same
residue after division by $N$. As we know that there exists solution
to this system (\ref{qcond}) with  $a_0=1$,
we conclude that there exists solution where the only nonzero
coefficients are
$$
a_0, a_N, a_{2N}, \ldots, a_{(m+n+\delta)N},
$$
where $\delta = 0$ if $1\le i \le N-1$, and $\delta=1$ if $N+1\le i \le
2N-1$. The corresponding quasi-invariant has the form
\beq{qsplit}
R^1_i= \sum_{s=0}^{m+n+\delta} a_{sN}
z^{(m+n-s)N+i}\bar z^{Ns}
\eeq
with the coefficients satisfying the following system of equations
$$
 \sum_{\nad{0\le s\le m+n+\delta}{s=2k}} a_{sN}
((m+n-2s)N+i)^{2t-1}=0, \quad 1\le t \le n,
$$
\beq{qsplitcond}
 \sum_{\nad{0\le s\le m+n+\delta}{s=2k+1}} a_{sN}
((m+n-2s)N+i)^{2t-1}=0, \quad 1\le t \le n,
\eeq
$$
\sum_{s=0}^{m+n+\delta} a_{sN}
((m+n-2s)N+i)^{2t-1}=0, \quad n< t \le m.
$$
Since the quasi-invariance equations (\ref{qcond} (or
\ref{qsplitcond}))
are real we conclude that
complex conjugate to a quasi-invariant would also be a
quasi-invariant. Therefore there exist quasi-invariants
$$
R^2_i= \bar R^1_i= \sum_{s=0}^{m+n+\delta}b_{sN}
\bar z^{(m+n-s)N+i} z^{Ns}
$$
with $b_{sN}=\bar a_{sN}$.
For $i\le N-1$ we have $\delta=0$ and we may define $q^1_i=R^1_i$. Then
$q^1_i$ have a form required by the statement of the Proposition.
For $i\ge N+1$ the last nonzero coefficient in $R^1_i$ is
$a_{(m+n+1)N}$. Let us show that one can make $a_{(m+n+1)N}=0$.
For that consider quasi-invariant
$$
q_i^1=R_i^1- a_{(m+n+1)N} (z\bar z)^{i-N}R^2_{2N-i}.
$$
We have
$$
(z\bar z)^{i-N}R^2_{2N-i} = (z\bar z)^{i-N} \sum_{s=0}^{m+n} b_{sN}
\bar z^{(m+n-s)N+2N-i} z^{Ns} =
$$
$$
\sum_{s=0}^{m+n} b_{sN}
\bar z^{(m+n-s+1)N} z^{Ns+i-N}=
\sum_{\tilde s=0}^{m+n} b_{(m+n-\tilde s)N}
z^{(m+n-\tilde s)N+i-N} \bar z^{N\tilde s+N}.
$$
Therefore
$$
q^1_i = z^{(m+n)N+i} + \sum_{s=1}^{m+n+1} z^{(m+n-s)N+i} \bar
z^{Ns} (a_{sN}-a_{(m+n+1)N} b_{m+n-s+1})
$$
and the last term in the sum above is equal to zero since $b_0=1$.
Thus $q^1_i$ for $i\ge N+1$ also have a form required in the
Proposition. Taking complex conjugates we get the quasi-invariants of the form
$$
q^2_i = \sum_{s=0}^{m+n} b_{s}\bar z^{(m+n-s)N+i} z^{Ns},  \quad
b_{s}=\bar a_{s}.
$$
The Proposition is proved.
\begin{theorem}\lb{freebasisT}
The quasi-invariants $q_i^{1,2}$ together with $q^0, q^1, q^2, q^3$
form a free basis for $Q^{I_2(2N)}$ over $S^{I_2(2N)}$.
\end{theorem}
By Theorem \ref{EGt} a set of vectors forms a basis for $Q^{I_2(2N)}$ over $S^{I_2(2N)}$
if it forms a basis in the complement to the ideal $I$. By
Proposition \ref{freebasis} the polynomials $q^0, q^1, q^2, q^3$
form a part of a basis for $Q^{I_2(2N)}$ over invariants. We are
left to prove that  constructed in Proposition \ref{freebasis} quasi-invariants do not
belong to the ideal $I$ generated by homogeneous invariants of positive degree. More exactly
one should prove that the linear space $\langle q^1_i,
q^2_i\rangle \cap I = 0$.
If the intersection is not trivial then we would have that
$$
\lambda_1 q^1_i + \lambda_2 q^2_i \in I,
$$
and
$$
\lambda_1 z^{(m+n)N+i} + \lambda_2 \bar z^{(m+n)N+i} + O(z\bar z)
= p(z,\bar z) \in I,
$$
where $O(z \bar z)$ is a polynomial divisible by $z \bar z$, and
 $\lambda_1, \lambda_2$ are  some constants.
Now, in the right-hand side we should have $p(z,\bar z)$ as a linear combination of
basis vectors for the complement to $I$ multiplied by polynomial
coefficients in $z\bar z$ and $z^{2N}+\bar z^{2N}$. The degrees of
basis vectors are given in the table on page \pageref{table}. We
see that there are no basis vectors whose degree would be less than
$(m+n)N+i$ but equal to $(m+n)N+i$ modulo $2N$. Therefore $p(z,
\bar z)$ must be divisible by $z\bar z$, hence
$$
\lambda_1 z^{(m+n)N+i} + \lambda_2 \bar z^{(m+n)N+i} \quad \mbox{ is
divisible by }\, z\bar z
$$
which is possible only if $\lambda_1=\lambda_2=0$. Thus the theorem is
proven.

\section{Explicit formulas for $\pmb m$-harmonic basis}

Let us show that the basis for $Q^{I_2(2N)}$ just considered in the
theorem is actually $m$-harmonic. Let us denote the quantum
integrals corresponding to the invariants $z \bar z$ and
$z^{2N}+\bar z^{2N}$ by ${\cal L}_1$ and ${\cal L}_2$
respectively. The corresponding operators have the forms (c.f.
\cite{FV2})
\beq{L1}
{\cal L}_1 = 4 \p_z \p_{\bar z}-4m\sum_{\nad{k=0}{k=2j}}^{2N-1}
\frac{e^{i\varphi_k}\partial_z-e^{-i\varphi_k}\partial_{\bar
z}}
{-e^{-i\varphi_k}z+e^{i\varphi_k}\bar z}
-4n \sum_{\nad{k=1}{k=2j+1}}^{2N-1}
\frac{e^{i\varphi_k}\partial_z-e^{-i\varphi_k}\partial_{\bar
z}}
{-e^{-i\varphi_k}z+e^{i\varphi_k}\bar z}
\eeq
where $\vf_k=\frac{\pi k}{2N}$. As to operator ${\cal L}_2$ it has
the following form (up to a constant)
$$
{\cal L}_2=\p_z^{2N}+\p_{\bar z}^{2N}+ \mbox{ lower order terms. }
$$

\begin{theorem}\lb{dihargen}
Polynomials $(\ref{q}), (\ref{bazis})$ are $m$-harmonic, that is,
the following identities hold
\begin{enumerate}
\item
${\cal L}_1(q^s) = {\cal L}_2(q^s) = 0, \qquad 0\le s \le 3$.
\item
${\cal L}_1(q_i^s) = {\cal L}_2(q_i^s) = 0, \qquad 1\le i \le 2N-1,\, i\ne N,\,
s=1,2.$
\end{enumerate}
\end{theorem}

\noindent
{\bf Proof} Let us first establish that polynomials $q^s$ are
$m$-harmonic. We have ${\cal L}_1(1)={\cal L}_2(1)=0$ as $\deg {\cal
L}_i(1)<0$ but on the other hand ${\cal L}_i(Q^{I_2(2N)})\subset
Q^{I_2(2N)}$ (see \cite{FV1} where this is explained in general case),
that is the image is polynomial. Further let us prove that quasi-invariant $q^3$ is $m$-harmonic.
Notice that
$$
q^3=\mu_3 \prod_{i=0}^{2N-1}(\a_i,x)^{2m_i+1}
$$
is characterized by the property that it is anti-invariant
quasi-invariant for the group $I_2(2N)$ of minimal possible
degree. Since operators ${\cal L}_1, {\cal L}_2$ are
$I_2(2N)$-invariant and since their degrees are negative we get
that
$$
{\cal L}_1(q^3)={\cal L}_2(q^3)=0.
$$
Quasi-invariants $q^1, q^2$ are anti-invariant quasi-invariants of
minimal possible degree with respect to subgroups $I_2(N)\subset
I_2(2N)$ corresponding to two different orbits of mirrors of the
system $I_2(2N)$. Since  ${\cal L}_1$, ${\cal L}_2$ are invariants with
respect to these subgroups we get the rest equalities of the first
statement of the theorem, namely
$$
{\cal L}_1 (q^{1,2})={\cal L}_2 (q^{1,2})=0.
$$
Let us now prove the second part. Consider two-dimensional space
$V=\langle q_i^1, q_i^2\rangle$ generated by polynomials $q^1_i,
q^2_i$. Let us introduce two-dimensional spaces $U,W$ generated by
the image of $V$ under the action of operators ${\cal L}_1 , {\cal
L}_2$, namely
$$
U=\langle{\cal L}_1 q_i^1, {\cal L}_1 q_i^2\rangle,
$$
$$
W=\langle{\cal L}_2 q_i^1, {\cal L}_2 q_i^2\rangle.
$$
As operators ${\cal L}_1, {\cal L}_2$ are invariant with respect
to $I_2(2N)$ action, they are intertwining operators between
representations $V$ and $U$, and between $V$ and $W$
correspondingly. According to Schur lemma either these operators
are zero operators or one or two of them is scalar operator. The
first case corresponds to the conditions 2 in the statement of the
theorem. In the second case corresponding representation $U$ (or
$W$) is isomorphic to representation $V$.

Representation $V$ is two-dimensional irreducible representation.
In the basis $q_i^1, q_i^2$ generating matrices $s$ and $s\tau$
(see (\ref{sst})) take the form
\beq{mat}
s=\left(
\begin{array}{cc}
0&1\\
1&0
\end{array}
\right),\qquad
s\tau= (-1)^{m+n}
\left(\begin{array}{cc}
\epsilon^i&0\\
0&\epsilon^{-i}
\end{array}
\right).
\eeq
For any $ q\in W$ one has $\deg q =(m+n)N+i-2N<(m+n)N.$
Since $Q^{I_2(2N)}$ is freely generated by $q^i, q_i^{1,2}$ over
$S^{I_2(2N)}$, the
two-dimensional irreducible representations in $Q^{I_2(2N)}$ are met in
the degrees more or equal to
$$
\min_i (\deg q_i^{1,2}) = (m+n)N+1.
$$
Hence $W=0$. Let us show that $U=0$ as well.
Notice that as it follows from (\ref{mat}) among representations
$V_i=\langle q_i^1, q_i^2\rangle$ each irreducible two-dimensional
representation of $I_2(2N)$ is met twice. Namely, $V_i\cong
V_{2N-i}$, and under this isomorphism
$$
%\begin{array}
q_i^1 \to \lambda q^2_{2N-i},
$$
$$
q_i^2 \to \lambda q^1_{2N-i}.
%\end{array}
$$
If $1\le i\le N-1$ then representation $V_i$ does not occur in the
space $Q^{I_2(2N)}$
in the degrees less than $i$. Therefore
$$
{\cal L}_1(q^1_i) = {\cal L}_1(q^2_i) = 0.
$$
We are left to consider the case $N+1\le i \le 2N-1$. Generally
speaking we have
$$
{\cal L}_1(q^1_i) = p_1 q_{2N-i}^2,
$$
$$
{\cal L}_1(q^2_i) = p_2 q_{2N-i}^1,
$$
where $p_1, p_2\in S^{I_2(2N)}$. Let us show that it must be $p_1=p_2=0$.
Recall that
$$
\begin{gathered}
q^1_i = \sum_{s=0}^{m+n} a_{s} z^{(m+n-s)N+i} \bar z^{Ns}, \quad
\\
q^2_{2N-i} =\sum_{s=0}^{m+n} b_{s}\bar z^{(m+n-s)N+2N-i} z^{Ns}.
\end{gathered}
$$
Notice that the degree of $q^1_i$ with respect to variable $\bar
z$ is equal to $(m+n)N$ at most. Because of the form (\ref{L1}) of
operator ${\cal L}_1$ we get
$$
\deg_{\bar z} {\cal L}_1(q^1_i)\le
(m+n)N.
$$
On the other hand if $p_2\ne 0$ we obtain
$$
\deg_{\bar z} (p_2 q^2_{2N-i}) \ge \deg_{\bar z} q^2_{2N-i} =
(m+n)N+2N-i>(m+n)N.
$$
The contradiction shows that $p_2=0$. Similarly $p_1=0$,
 and therefore polynomials
$q^1_i, q^2_i$ are $m$-harmonic. The theorem is proven.
\begin{cor}\label{unique}
Quasi-invariants $q^1_i, q^2_i$ of the form $(\ref{bazis})$ are uniquely
defined.
\end{cor}
Indeed, the linear space of $m$-harmonics is defined uniquely (see
\cite{FV1}). The space $H_i$ of homogeneous  $m$-harmonics of degree $(m+n)N+i$ is
two-dimensional. Therefore  quasi-invariants $q^1_i, q^2_i$ are uniquely characterized
as homogeneous $m$-harmonics  of the form
$$
z^{(m+n)N+i}+O(z\bar z), \quad \bar z^{(m+n)N+i}+O(z\bar z)
$$
correspondingly, where $O(z \bar z)$ denotes polynomials divisible
by $z \bar z$.

%%%%%%%%%%%%

Finally let us present explicit determinant formulas for
quasi-invariants $q^1_i$, $q^2_i$. Let us introduce (see
p.~\pageref{MatrixA}) the
$(m+n+1)\times (m+n+1)$ matrix $A = A(i)$.

\begin{landscape}
\null\vfill
\begin{scriptsize}
$$\label{MatrixA}
A=
\begin{pmatrix}
z^{(m+n)N+i} & \ol{z}^N z^{(m+n-1)N+i}
             & \ol{z}^{2N} z^{(m+n-2)N+i}
             & \multicolumn{4}{c}{\dotfill}
             & \ol{z}^{(m+n)N} z^i
%%%%%%% blok
\\[7mm]
(m+n)N+i & 0
         & (m+n-4)N+i
         & 0
         & (m+n-8)N+i
         & 0
         & \multicolumn{2}{c}{\dotfill}
\\[2mm]
((m+n)N+i)^3 & 0
             & ((m+n-4)N+i)^3
             & 0
             & ((m+n-8)N+i)^3
             & 0
             & \multicolumn{2}{c}{\dotfill}
\\[2mm]
\multicolumn{8}{c}{\dotfill}
\\[2mm]
((m+n)N+i)^{2n-1} & 0
             & ((m+n-4)N+i)^{2n-1}
             & 0
             & ((m+n-8)N+i)^{2n-1}
             & 0
             & \dots
             & \dotfill
\\[7mm]
0 & (m+n-2)N+i
  & 0
  & (m+n-6)N+i
  & 0
  & \multicolumn{3}{c}{\dotfill}
\\[2mm]
0 & ((m+n-2)N+i)^3
  & 0
  & ((m+n-6)N+i)^3
  & 0
  & \multicolumn{3}{c}{\dotfill}
\\[2mm]
 \multicolumn{8}{c}{\dotfill}
\\[2mm]
0 & ((m+n-2)N+i)^{2n-1}
  & 0
  & ((m+n-6)N+i)^{2n-1}
  & 0
  &  \multicolumn{3}{c}{\dotfill}
%%%%%%%%%%% blok
\\[7mm]
((m+n)N+i)^{2n+1} & ((m+n-2)N+i)^{2n+1}
                 & ((m+n-4)N+i)^{2n+1}
                 & ((m+n-6)N+i)^{2n+1}
                 & ((m+n-8)N+i)^{2n+1}
                 &  \multicolumn{3}{c}{\dotfill}
\\[2mm]
 \multicolumn{8}{c}{\dotfill}
\\[2mm]
((m+n)N+i)^{2m-1} & ((m+n-2)N+i)^{2m-1}
             & ((m+n-4)N+i)^{2m-1}
             & ((m+n-6)N+i)^{2m-1}
             & ((m+n-8)N+i)^{2m-1}
             &  \multicolumn{3}{c}{\dotfill}
\end{pmatrix}
$$
\end{scriptsize}
\vfill\null
\end{landscape}

\begin{theorem}\label{formulas}
For generators $q^1_i, q^2_i (1\le i\le 2N-1, i\ne N)$ the
following formulas hold
\beq{forms}
q^1_i=c \det A, \quad  q^2_i=c \det \bar A,
\eeq
where $c=(\det A_1)^{-1}$, and the matrix $A_1$
is obtained from $A$ by deleting the first column and the first
row.
\end{theorem}

\noindent
{\bf Proof}
Firstly let us show that $\det A_1\ne 0$.
During the proof of Theorem \ref{freebasisT}
we showed (see formulas (\ref{qsplit}), (\ref{qsplitcond})) that quasi-invariance equations
on the coefficients
of polynomial
$$
q^1_i = \sum_{s=0}^{m+n} a_{s} z^{(m+n-s)N+i} \bar z^{Ns}, \quad
a_0=1
$$
have the form
$$
 \sum_{\nad{0\le s\le m+n}{s=2k}} a_{s}
((m+n-2s)N+i)^{2t-1}=0, \quad 1\le t \le n,
$$
\beq{quasisyst}
 \sum_{\nad{0\le s\le m+n}{s=2k+1}} a_{s}
((m+n-2s)N+i)^{2t-1}=0, \quad 1\le t \le n,
\eeq
$$
\sum_{s=0}^{m+n} a_{s}
((m+n-2s)N+i)^{2t-1}=0, \quad n< t \le m.
$$
We know by theorem \ref{freebasisT} that this system has a solution. Condition
$\det A_1\ne 0$
is equivalent to the statement that system (\ref{quasisyst}) has a unique
solution since the matrix of this linear system is exactly $A_1$.
But the solution is unique according to Corollary \ref{unique}.

To prove the theorem we need to show that
\beq{kramer}
(-1)^{k+1}\frac{\det A_k}{\det A_1} = a_{k-1},
\eeq
$k=1, \ldots, m+n+1$,
where the matrix $A_k$ is obtained from $A$ by deleting the $k$th column and the first row.
According to Kramer's rule the solution $a_1,
\ldots, a_{m+n}$ to linear system (\ref{quasisyst}) is given by the formulas
$$
a_{k-1} = \frac{\det \hat A_{k}}{\det A_1},
$$
where $\hat A_{k}$ is  matrix $A$ where the first column and the first row are
omitted, and instead of $k$th column it is written the first column
with a minus sign. Changing the sign for this column and interchanging
it to the first place we get
$$
\det \hat A_{k} = (-1)(-1)^{k-2}\det A_k,
$$
thus (\ref{kramer}) is satisfied. Theorem is proven.

\vspace{7mm}
\noindent
{\bf Remark.} We have deduced formulas for $q_i^{1,2}$ under
assumption that $n<m$. Nevertheless Theorem \ref{formulas} remains
true and gives formulas for the basis of quasi-invariant module consisting of
$m$-harmonic polynomials also in the case $m=n$. One can see that in this case
formulas \mref{forms} reduce to the formulas for $m$-harmonic
basis obtained in \cite{FV2}.

%%%%%%%%%%%%%%%%%%%%%%%%%%%%%%%%%

\section{Concluding remarks}

\subsection{Generators for general Coxeter systems}

In this paper it is completed the description of quasi-invariant
modules for two-dimensional Coxeter systems. The description of
generators of $Q^\cR$ over $S^G$ in the dimensions greater than
two is an open question. In the general case the choice of
$m$-harmonic polynomials is impossible as it is shown by Etingof
and Ginzburg \cite{EG}.
But the counterexample found by the authors looks
exceptional as it consists of the Coxeter system $C_6$
with multiplicities on the long roots equal 0. The ring of
corresponding quasi-invariants may be described as a module over
invariants of the Coxeter group $A_1^6$ with the generators which are
$m$-harmonic polynomials related to the group $A_1^6$.
The question when $m$-harmonic polynomials can be chosen as the
generators of $Q^\cR$ over $S^G$ is an open question when the
dimension is greater than two.

\subsection{Deformed systems}
In the papers \cite{FV3,CFV3}
the quantum systems of Calogero--Moser type related to non-Coxeter
configurations $\cA=\anm, \cnm$ were investigated. The
corresponding second order operator has the form
$$
\cl^\cA=\Delta-\sum_{\a\in\cA}\frac{2m_\a}{(\a,x)}\p_\a.
$$
As in the Coxeter case the commutative ring of quantum integrals
for $\cl^\cA$ is isomorphic to the ring $Q^\cA$ of quasi-invariants
related to the system $\cA$.
It is shown in \cite{FV3} that the rings $Q^\cA$ are
Cohen--Macaulay, that is they are freely generated over polynomial
subrings $S^\cA$, which are analogs of the invariant rings in the
case of Coxeter systems. The question of description
of the rings
$Q^\cA$, and in particular the question on the generators of $Q^\cA$
as a module over
$S^\cA$ is an open problem.
The natural generalization of $m$-harmonic polynomials to the
non-Coxeter case does not lead to a basis already in the simplest case
of the system ${\cal A}_2(2)$ (see \cite{FV3}).

\vspace{10mm}

\noindent
{\bf Acknowledgements.} I would like to thank A.P.Veselov for
useful and stimulating discussions. I am grateful to N.Andreev for his help.
On the final stage the work was supported by Russian President grant MK-2050.2003.01.

\noindent Department of Mathematics, South Kensington campus, Imperial College, London, SW7 2AZ, UK. \\
email: m.feigin@imperial.ac.uk

\end{document}